%Paper: hep-th/9404171
%From: Aida <aida@phys.titech.ac.jp>
%Date: Thu, 28 Apr 1994 17:57:18 +0900

\input phyzzx
\FRONTPAGE
\line{\hfill TIT-HEP-256}
\line{\hfill KEK-TH-395}
\line{\hfill YITP/U-94-13}
\line{\hfill April 1994}
\vskip .15in
%\vskip 1.5truein
\titlestyle{{Conformal Invariance and Renormalization Group in Quantum Gravity
Near Two Dimensions}}
\author{Toshiaki Aida, Yoshihisa Kitazawa}
\centerline{{\it  Department of Physics}}
\centerline{{Tokyo
Institute of Technology,}}
\centerline{Oh-okayama Meguroku Tokyo 152, Japan}
%\vskip .15in
\author{Hikaru Kawai}
\centerline{{\it National Laboratory for High Energy Physics (KEK)}}
\centerline{{ Tukuba, Ibaraki 305, Japan}}
%\bigskip
%\vskip .15in
\centerline{{\it and}}
%\vskip .15in
\author{Masao Ninomiya}
\centerline{{\it Uji Research Center, Yukawa Institute for Theoretical
Physics}}
\centerline{ Kyoto University, Uji 611, Japan}
\bigskip

\abstract
We study quantum gravity in $2+\epsilon$ dimensions
in such a way to preserve the volume
preserving diffeomorphism invariance. In such a formulation,
we prove the following trinity: the general covariance, the conformal
invariance and the renormalization group flow to Einstein theory
at long distance.
%We investigate renormalization of quantum fields to study higher
%loop corrections.
We emphasize that the consistent and macroscopic universes
like our own can only
exist for matter central charge $0<c<25$.
We show that the spacetime singularity at the big bang is resolved by
the renormalization effect and universes are found to bounce back
from the big crunch.
Our formulation may be viewed as a Ginzburg-Landau theory
which can describe both the broken and the unbroken phase
of quantum gravity and the phase transition between them.

\endpage

\centerline{{\bf 1.  Introduction}}

The reconciliation of quantum mechanics and Einstein's theory of gravitation
has been a long standing problem in theoretical physics.
It has lead to the modification of Einstein's general theory of relativity
such as string theory.
However in two dimensions, this conflict can be resolved because Einstein
gravity is power counting renormalizable.
The remarkable point is the asymptotic freedom of the gravitational
coupling constant near two dimensions. Beyond two dimensions, the gravitational
coupling constant acquires the canonical dimension. So it effectively grows
at short distance. However this growth is counter balanced by the
asymptotic freedom and consistent quantum theories may be constructed
even beyond two dimensions. $2+\epsilon$ expansion of quantum gravity is base
on
such an idea[1][2][3]. In view of the exact solution of two dimensional
gravity[4], this approach may be developed further[5].

Although this idea works near two dimensions, the draw back is that
we have to put $\epsilon =1$ or $2$ to reach three or four dimensions.
One of the motivations to pursue this approach comes from the recent progress
in dynamical triangulation method
in realistic dimensions. The existence of phase transitions
itself is encouraging and the strong coupling phase appears to resemble two
dimensional quantum gravity[6]. However physically interesting weak coupling
region is afflicted with the conformal mode instability and may require
great care to extract physics.

Another motivation to study quantum gravity in $2+\epsilon$ dimensions is
to regard it as a toy model and try to learn lessons for realistic
quantum gravity. In this regard we can cite the questions concerning spacetime
singularities which appear at the big bang or the end of the blackhole
evaporation.
Of course various questions which arise from the existence of event horizons
are also interesting to analyze in this context.
We can even think of string theory application of this approach by taking
$\epsilon \rightarrow
0$ limit.

We use dimensional regularization to compute Feynman diagrams which respects
the crucial symmetries of the theory, namely general covariance and
conformal invariance.
In quantum gravity, the metric $g_{\mu\nu}$ is the fundamental dynamical
quantity to
integrate. We decompose the metric into the conformal mode $\phi$ and the
traceless symmetric
tensor $h_{\mu\nu}$.
%$h_{\mu\nu}$ as $g_{\mu\nu} = (e^h)_{\mu\nu} e^{-\phi} $.
It is also convenient to introduce a reference metric $\hat{g}_{\mu\nu}$.
We parametrize the metric of the space-time as $g_{\mu\nu} =
\tilde{g}_{\mu\nu}e^{-\phi} = \hat{g}_{\mu\rho} {(e^h )^{\rho}}_{\nu}
e^{-\phi}$
where
${h^{\mu}}_{\nu}$ field is
taken to be traceless ${h^{\mu}}_{\mu} =0$, while $h_{\mu\nu}$ is symmetric
in $\mu$ and $\nu$. The tensor indices are raised and lowered by the background
metric $\hat{g}_{\mu\nu}$.
%where $h^{\mu}_{\mu} =0$.

The dynamics of these two
fields turn out to be very different.
We need the Einstein action as a counter term to cancel the divergences
of correlation functions of $h_{\mu\nu}$ field.
However such a counter term is an oversubtraction
for the conformal mode. This situation is easy to understand in conformal
gravity since the one loop amplitudes do not involve the conformal mode.
The Einstein gravity becomes conformally invariant in two dimensions.

The bare Einstein action of the theory which contains $c$ copies of
real scalar fields
in our parametrization is
$$\eqalign{
{1\over G^0} &\int d^D x \sqrt{g} R \cr
=\mu ^{\epsilon} ({1\over G} -{25-c\over 24\pi\epsilon})
&\int d^D x\sqrt{\hat g}e^{-{\epsilon\over 2}\phi}\{\tilde{R} -
{\epsilon (D-1)\over 4}
\partial _{\mu}\phi\partial _{\nu}\phi\tilde{g}^{\mu\nu}\} \cr
\rightarrow  {Q^2_{eff} \over 32\pi}
&\int d^D x \sqrt{\hat g} (\partial _{\mu}\phi
\partial ^{\mu}\phi +2\phi \hat R) .
}\eqno\eq$$
In this expression, $G^0$ and $G$ are the bare and the renormalized
gravitational
coupling constant respectively. $\mu$ is the renormalization scale to define
$G$.
The effective central charge $(c_{eff})$ which governs the dynamics
of the conformal mode can be defined as
$Q^2_{eff} = {25-c_{eff}\over 3}={25-c\over 3} -{8\pi\epsilon \over G}$.
Hence
the Liouville theory is obtained for $\phi$ field
and the exact gravitational scaling exponents are reproduced
by taking $\epsilon \rightarrow 0$ limit in the strong coupling regime.
The novel feature beyond two dimensions is that the gravitational coupling
constant
acts as the effective central charge for the dynamics of the conformal mode
$\phi$[7][8].
This approach is further studied in [10][11].
The work which addresses relevant questions to ours is [9].

However the renormalization of $\phi$ field is
hard to keep track in a manifestly generally covariant form. We have proposed
to formulate
the theory in such a way to preserve the manifest volume preserving
diffeomorphism
invariance[8]. In such a formalism, the general covariance can be recovered by
further
demanding the conformal invariance on the theory. Another idea to ensure the
general covariance
on the theory is to consider the renormalization group trajectory in the
coupling constant
space which leads to the Einstein gravity at long distance.
In this paper we study the effectiveness of these ideas in detail by analyzing
conformally
coupled Einstein gravity.
%For theories which preserve the manifest volume preserving
%diffeomorphism, we prove the validity of our proposal.
We prove the validity of our proposal to the one loop order and give a concrete
prescription to calculate higher order corrections.

It is often thought that we may be in the broken phase of quantum gravity
since the expectation value of the metric is certainly nonvanishing
in our Universe. We may think of the unbroken phase of quantum gravity
in which the metric is fluctuating around the vanishing expectation value.
As it turns out that our formalism can describe not only the broken phase of
quantum gravity but also the unbroken phase . Furthermore it can study the
phase transition between them. In the phase transition theory
of statistical systems, the Ginzburg-
Landau theory plays an important conceptual role to discuss symmetry breaking
and the order of phase transitions and so on.
In quantum gravity, it is certainly desirable to have such a description.
Our formalism supplies such a conceptually crucial tool to investigate
quantum gravity.

We consider Minkowski metric in this paper although Euclidian rotation
of the theory is valid at the Feynman perturbation theory level.
We perform real calculations in that way. However Euclidian quantum gravity
beyond two dimensions is unstable due to the negative sign of the kinetic
term of the conformal mode in the weak coupling regime.
Presumably such a theory does not possess
a stable vacuum. This feature is what we expect for quantum gravity because
the Universe has begun with the big bang and it is still expanding.

The organization of this paper is as follows. In section 2, we consider the
conformally coupled
Einstein gravity. In section 3, we renormalize the theory to the one loop level
and
the renormalization
group equations are derived. In section 4, we prove the trinity in our
formalism: the general
covariance, the conformal invariance and the renormalization group flow to the
Einstein gravity
at long distance. In section 5, we study the subtractions of the internal loops
in preparation for
the two loop renormalization of the theory. In section 6, we study the
renormalization of the
cosmological constant operator. In section 7, we consider the cosmological
implications of this
model. The last section consists of the conclusions and discussions.

\vskip .10in

\centerline{{\bf 2.  Conformal Gravity in $2+\epsilon$ Dimensions}}

\vskip .10in

Let us couple a real scalar field $\psi$ to Einstein gravity in
a conformally invariant way:
$$ \int d^D x\sqrt{g}\{R{\epsilon\over {8(D-1)}}\psi ^2 -
{1\over 2} \partial _{\mu}\psi\partial _{\nu}\psi g^{\mu\nu}\} .
\eqno\eq $$
%Namely we decompose the
%metric into the conformal factor and the traceless symmetric tensor
%$h_{\mu\nu}$.
Since the action is conformally invariant, the conformal mode dependence
drops out of the action as,
$$ \int d^D x \sqrt{\hat g}\{\tilde{R}{\epsilon\over {8(D-1)}}\psi ^2 -
{1\over 2} \partial _{\mu}\psi\partial _{\nu}\psi\tilde{g}^{\mu\nu}\} ,
\eqno\eq $$
where $\tilde{R}$ is the scalar curvature made out of $\tilde{g} _{\mu\nu}$.
By changing the variable as $\sqrt{{\epsilon}\over {8(D-1)}}\psi =
e^{-{{\epsilon}\over{4}} \phi}$, we obtain
$$ \int d^D x \sqrt{\hat g}e^{-{\epsilon\over 2}\phi}\{\tilde{R} -
{\epsilon (D-1)\over 4}
\partial _{\mu}\phi\partial _{\nu}\phi\tilde{g}^{\mu\nu}\} .
\eqno\eq $$

Note that this is nothing but the Einstein action. Therefore we can regard
the conformal mode as a matter field which couples to the Einstein gravity
in a conformally invariant way. The Einstein gravity can be quantized from
such a view point as long as we maintain the conformal invariance.
In fact this strategy is that of D.D.K. in two dimensional quantum gravity
[12][13].
We have proposed to generalize this strategy (or nonlinear sigma model
approach to string theory) into $2+\epsilon$ dimensions[8].

However we point out the crucial difference between (3) and (4).
Namely $\psi =0$ point is mapped to $\phi =\infty$ during the change
of the variables. The metric $g_{\mu\nu}$ vanishes at this point.
Therefore it is impossible to study the fluctuations of the metric around such
a point
in the manifestly generally covariant form (4).
However in (3), the expansion around $\psi=0$ point does not
present any problems except that the $\tilde {R}$ decouples at this point.
As it turns out, this problem can be overcome in the quantum formulation due
to the one loop quantum effect.
Therefore our formulation has the definite advantage of being capable to
study the unbroken phase of quantum gravity in which the metric is fluctuating
around the vanishing expectation value.

In our previous work, we have studied Einstein gravity with minimally coupled
scalar fields. In this paper we consider even simpler models, namely
Einstein gravity with $c$ copies of conformally coupled scalar fields.
The action we consider is,
$$\eqalign{S = {\mu ^{\epsilon} \over G}
\int d^D x\sqrt{g}\{&{R}(1-{\epsilon\over 8(D-1)}\varphi _i^2) +
{1\over 2} \partial _{\mu}\varphi _i\partial _{\nu}\varphi _i
{g^{\mu\nu}}\} \cr
 = {\mu ^{\epsilon} \over G}
 \int d^D x \sqrt{\hat g}\{&\tilde{R}((1+{1\over 2}\sqrt{{\epsilon\over
2(D-1)}}\psi )^2
-{\epsilon\over 8(D-1)}\varphi _i^2) \cr
 &- {1\over 2} \partial _{\mu}\psi\partial _{\nu}\psi\tilde{g}^{\mu\nu}
+ {1\over 2} \partial _{\mu}\varphi _i\partial _{\nu}\varphi _i
\tilde{g}^{\mu\nu}\}.\cr}
\eqno\eq $$
We expand $\psi$ around the constant mode in (3):
$\sqrt{{\epsilon}\over {8(D-1)}}\psi \rightarrow
\tilde{\mu}^{\epsilon \over 4} (1+{1\over 2}\sqrt{{\epsilon\over 2(D-1)}}\psi
)$.
We choose the renormalization scale $\mu$ to compensate the scale
$(\tilde{\mu})$ of
$\psi$ and $\varphi_{i}$ fields. By doing so, we obtain the above action.
Therefore in order to probe the beginning of the universe where $\psi$ is
small,
we need to consider the large renormalization scale $\mu$.

Our strategy is to impose the conformal invariance on the theory
while regarding the conformal mode as a matter field.
However the conformal invariance is in general broken by quantum effects.
Hence we have to cancel the quantum conformal anomaly by the tree level
conformal anomaly.
For this purpose, we
have proposed to generalize the Einstein gravity as
$$ {\mu ^{\epsilon} \over G}
\int d^D x \sqrt{\hat g}\{\tilde{R}L(\psi ,\varphi _i )
-{1\over 2} \partial _{\mu}\psi \partial _{\nu}\psi
{\tilde{g}^{\mu\nu}}
+{1\over 2} \partial _{\mu}\varphi _i\partial _{\nu}\varphi _i
{\tilde{g}^{\mu\nu}}\} .
\eqno\eq $$
where $L(\psi ,\varphi _i )$ is a function of $\psi$ and $\varphi _i$ with
the constraint $L(0,0)=1$.
Note that in conformally coupled Einstein gravity,
$$ L(\psi ,\varphi _i )=1+\sqrt{{\epsilon\over {2(D-1)}}}\psi
+{\epsilon\over {8(D-1)}}(\psi ^2 - \varphi _i^2 ) .
\eqno\eq $$

Since we have generalized the Einstein gravity, the tree action is
no longer generally covariant. However our action is still invariant
under
the volume preserving diffeomorphism.
%The difference between these
%transformations is the conformal transformation.
We consider the following gauge transformation
of the fields,
$$\eqalign{\delta \tilde{g}_{\mu\nu}&=
\partial _{\mu} \epsilon ^{\rho}\tilde{g} _{\rho\nu}
+\tilde{g} _{\mu\rho} \partial _{\nu}\epsilon ^{\rho}
+\epsilon ^{\rho}\partial _{\rho}\tilde{g} _{\mu\nu}
-{2\over D}\partial _{\rho} \epsilon ^{\rho} \tilde{g}_{\mu\nu} ,\cr
\delta \psi &= \epsilon ^{\rho} \partial _{\rho} \psi
+ (D-1) {\partial L\over {\partial \psi}}{2\over D}\partial _{\rho}
\epsilon ^{\rho} ,\cr
\delta \varphi _i &= \epsilon ^{\rho} \partial _{\rho} \varphi _i
- (D-1) {\partial L\over {\partial \varphi _i}}{2\over D}\partial _{\rho}
\epsilon ^{\rho} .\cr} \eqno\eq$$
If $L$ is given by (7), this gauge invariance is nothing but the
general covariance.
The last term of each transformation can be viewed as a part of
a conformal transformation.
Since the action is invariant under the volume preserving diffeomorphism,
it is invariant under the gauge transformation (8) if it is invariant
under the following conformal transformation with respect to the background
metric $\hat{g}_{\mu\nu}$
$$\eqalign{
\delta \hat{g}_{\mu\nu} &= -\hat{g}_{\mu\nu} \delta \rho ,\cr
\delta h_{\mu\nu} &= 0 ,\cr
\delta \psi &= (D-1) {\partial L\over \partial \psi}\delta \rho ,\cr
\delta \varphi _i &= -(D-1) {\partial L\over \partial \varphi _i} \delta \rho .
}\eqno\eq$$
%Note that the action (5) is invariant
%under this gauge transformation
%if it is conformally invariant.
%Hence it is invariant under the volume preserving diffeomorphism
%where $\partial _{\mu} \epsilon ^{\mu} =0$.

Therefore the change of the action under this gauge transformation
is nothing but the conformal anomaly of the theory:
$$\eqalign{ {\mu ^{\epsilon} \over G}
\int d^D x\sqrt{\hat g} \{ & {1\over 2}\{\epsilon L - 2(D-1)(
({\partial L\over{\partial \psi}})^2
-({\partial L\over{\partial \varphi _i}})^2)\}\tilde{R} \cr
&-{1\over 4}\{\epsilon - 4(D-1){\partial ^2 L\over{\partial \psi ^2}}\}
\partial _{\mu}\psi\partial _{\nu}\psi\tilde{g}^{\mu\nu} \cr
&+{1\over 4}\{\epsilon +4(D-1){\partial ^2 L\over{\partial \varphi _i^2}}\}
\partial _{\mu}\varphi _i\partial _{\nu}\varphi _i\tilde{g}^{\mu\nu} \}
\delta \rho ,
}\eqno\eq $$
where $\delta \rho = {2\over D}\partial _{\rho} \epsilon ^{\rho}$.
Furthermore it vanishes for the Einstein gravity. Thus at the tree level,
a general covariant theory is obtained from a theory which is invariant
under volume preserving diffeomorphism by further demanding the conformal
invariance.  The advantage of this strategy is that it
works at the quantum level.
We have proposed to generalize the Einstein gravity to the most general
renormalizable form which only possesses the volume preserving
diffeomorphism invariance. The general covariance can be recovered by further
demanding the conformal invariance. Alternatively we may follow the
renormalization group trajectory of the most general renormalizable
theory
which leads to the Einstein gravity
in the weak coupling limit. In this paper we investigate the effectiveness
of these ideas in detail.
% and
%prove the equivalence of these approaches
%at the one loop level.

\vskip .10in

\centerline{{\bf 3.  1-Loop Renormalization and Renormalization Group}}

\vskip .10in

In this section we compute the one loop counter terms and derive the
renormalization group equations for the coupling constants
in conformally coupled Einstein gravity (6).
We expand the fields around the classical part(backgrounds) and
employ a background gauge.
In the background gauge the manifest volume preserving diffeomorphism
invariance is maintained.
This procedure has been explained in [7][8].
In such a scheme,
$\tilde{g} _{\mu\nu}$ is expanded around the background metric
$\hat{g}_{\mu\nu}$ as $\tilde{g}_{\mu\nu} =\hat{g}_{\mu\rho}
{(e^{h})^{\rho}}_{\nu}$.
We would like to keep the gauge invariance of (8)
as the symmetry of the theory.

We adopt the following gauge fixing term,
$$
%\int d^D x \sqrt{\hat g}
{1\over 2}L(\bar\psi +\psi ,\bar\varphi _i +\varphi _i )
(\nabla ^{\mu}h_{\mu\nu} - {\partial _{\nu}L\over L})
(\nabla _{\rho}h^{\rho\nu} - {\partial ^{\nu}L\over L}),
\eqno\eq$$
where the covariant derivative is always taken with respect to the
background metric $\hat{g}_{\mu\nu}$.
The ghost action follows from the gauge fixing term in a standard way as
$$
%\int d^D x \sqrt{\hat g}
\nabla _{\mu} \bar \eta _{\nu} \nabla ^{\mu}\eta ^{\nu}
+{{\hat R}^{\mu}}_{\nu} {\bar \eta}_{\mu} \eta ^{\nu}
-{\partial _{\nu}L\over L}(\nabla ^{\mu}{\bar \eta}_{\mu})\eta ^{\nu}
+ \cdots .
\eqno\eq$$

The tree action is generalized such that it breaks this gauge invariance
at $O(\epsilon )$ because of the conformal anomaly.
The tree level conformal anomaly of $O({\epsilon \over G})$
is required to cancel the one
loop level conformal anomaly of $O(1)$ since $G$ is at most $O(\epsilon )$
around the short distance fixed point.
The important point is that
the one loop divergences of the theory have to respect this symmetry up to
finite terms since the symmetry is only `softly' broken at $O(\epsilon )$.
Therefore the one loop counter terms can be chosen to
possess the volume preserving
diffeomorphism invariance.
The exact gauge invariance can be recovered later by imposing the
conformal invariance on the bare action.

The one loop divergence of this theory is evaluated to be
$$
\int d^D x \sqrt{\hat g}
\{{26-(1+c)\over {24\pi \epsilon}}\}\tilde{R}.\eqno\eq
$$
Hence the one loop bare action is
$$
\int d^D x \sqrt{\hat g}
{\mu ^{\epsilon}\over G}\{\tilde{R}L(\psi ,\varphi _i ) -
{1\over 2}\partial _{\mu}\psi\partial _{\nu}\psi \tilde{g}^{\mu\nu}
+{1\over 2}\partial _{\mu}\varphi _i\partial _{\nu}\varphi _i
\tilde{g}^{\mu\nu} - {A\over \epsilon} G \tilde{R} \},\eqno\eq
$$
where $A={25-c\over {24\pi}}$.
We remark that a Dirac field gives rise to the same divergence
with a real conformally coupled scalar field. It is straightforward
to substitute a Dirac field for a conformally coupled scalar field.

In what follows, we derive the renormalization group equations for
the coupling constants. For this purpose, we define bare quantities as
$$
\eqalign{{1\over {G^0}} &= \mu ^{\epsilon}
({1\over G}-{A\over \epsilon}) ,\cr
{1\over {G^0}}\partial _{\mu} \psi _0 \partial ^{\mu} \psi _0
&= {\mu ^{\epsilon}\over {G}}\partial _{\mu} \psi \partial ^{\mu} \psi ,\cr
{1\over {G^0}}\partial _{\mu} \varphi _0^i \partial ^{\mu} \varphi _0^i
&= {\mu ^{\epsilon}\over {G}}\partial _{\mu} \varphi ^i
\partial ^{\mu} \varphi ^i ,\cr
L_0 (\psi _0 ,\varphi _0^i )
&= L(\psi ,\varphi ^i ) - {A\over \epsilon} G
+ {GA\over \epsilon} L(\psi ,\varphi ^i ). }
\eqno\eq$$
By solving these equations, we obtain
$$
\eqalign{\psi _0 &= \psi (1+{AG\over {2\epsilon}}) ,\cr
\varphi _0^i &= \varphi ^i (1+{AG\over {2\epsilon}}) ,\cr
L_0 (\psi _0 ,\varphi _0^i ) &= L(\psi _0,\varphi _0^i ) -
{AG\over \epsilon} + {AG\over \epsilon}L(\psi _0 ,\varphi _0^i ) \cr
&- {AG\over {2\epsilon}} \psi _0 {\partial L\over{\partial \psi _0}}
- {AG\over {2\epsilon}} \varphi _0^i {\partial L\over{\partial \varphi _0^i}} .
}\eqno\eq$$
$\beta$ and $\gamma$ functions (functionals to be precise)
follow by demanding that
the bare quantities do not
depend on the renormalization scale $\mu$ as
$$
\eqalign{\beta _G &=\epsilon G - AG^2 ,\cr
\beta _L &= -GA(L-1) + {GA\over 2}\psi {\partial L\over {\partial \psi}}
+ {GA\over 2}\varphi ^i {\partial L\over {\partial \varphi ^i}} ,\cr
\gamma _\psi &= -{AG\over 2} \psi ,\cr
\gamma _\varphi ^i &= -{AG\over 2} \varphi ^i .
}\eqno\eq$$

In this theory, $L$ contains only finite numbers of couplings.
Let us parametrize $L$:
$$
L=1+a\psi +b\psi ^2 -d\varphi _i^2 .
\eqno\eq$$
Then $\beta _L$ simplifies as
$$\eqalign{
\beta _a &= -{GA\over 2}a ,\cr
\beta _b &= 0 ,\cr
\beta _d &= 0 .
}\eqno\eq$$
So only two coupling constants ($G$ and $a$) possess nontrivial
$\beta$ functions in this theory. We further remark that we do not
need additional couplings to renormalize the theory to the one loop
order.

Since we have derived the renormalization group equations for the
couplings, we proceed to examine the renormalization group
trajectory of the theory
in the multidimensional space of the coupling constants.
In this theory, we need to consider the two dimensional coupling space of
$G$ and $a$.
At long distance where the renormalization scale $\mu$
is small, the gravitational coupling is also small. Under such
circumstances, the theory is classical and must agree with the
conformally coupled Einstein gravity. Therefore we are interested
in a particular renormalization group trajectory which starts with the
Einstein gravity in the weak coupling limit. We can trace such a
trajectory by solving $\beta$ functions.
It is clear by inspecting the $\beta$ functions that such a trajectory
is attracted to the ultraviolet stable fixed point at short
distance as long as $c<25$. At the fixed point
$G^* = {\epsilon\over A}$ and $a^* = 0$.
Therefore we conclude that the short distance structure of spacetime
is described by the following Lagrangian in conformal Einstein gravity near two
dimensions.
$$\eqalign{
\int d^D x \sqrt{\hat g} {\mu ^{\epsilon}\over {G^*}} &\{ \tilde{R}
(1+{\epsilon \over{8(D-1)}}(\psi ^2 - \varphi _i^2 ))\cr
-&{1\over 2}\partial _{\mu}\psi\partial _{\nu}\psi \tilde{g}^{\mu\nu}
+{1\over 2}\partial _{\mu}\varphi _i\partial _{\nu}
\varphi _i \tilde{g}^{\mu\nu}\} .
}\eqno\eq$$

The fixed point of the theory is realized at $\mu \rightarrow \infty$ limit.
This point corresponds to the vanishing expectation value of $\psi$
as it is explained in Sec. 2 and  in accord with the physical intuition.
As we have remarked in Sec. 2, $\tilde R$ does not decouple
at $\psi =0$ point due to the one loop quantum effect.
Furthermore the theory is still weak coupling at short distance since
$G^* \sim \epsilon$.
This theory is thus capable to describe the quantum fluctuations
of the metric around the vanishing expectation value.
We also note that we can rotate $\psi$ field into the complex direction
$\psi \rightarrow i\psi$ in the fixed point action while keeping the
reality of it.  By doing so, we obtain the Euclidian theory in which
the conformal mode becomes indistinguishable from the conformally
coupled matter fields.
As it is well known the short distance fixed point represents a
phase transition point.
In the opposite phase of quantum gravity beyond the short distance fixed point,
we may use such an Euclidian theory.
In the oversubtraction scheme of (1), we recall that
the sign of the kinetic term of the conformal mode flips
at the fixed point.  Our result is consistent with such a picture.

\vskip .10in

\centerline{{\bf 4. General Covariance, Conformal Invariance and
Renormalization Group}}

\vskip .10in

In this section, we would like to investigate the gauge invariance of
the theory in detail. In the weak coupling limit, the theory coincides
with conformally coupled Einstein gravity. It has the gauge invariance
which is nothing but the general covariance. The central question is
whether the general covariance is maintained along the renormalization
group trajectory. Since the renormalization group does not change the
physics, we expect that the answer to this question is affirmative.
In this section we prove that it is indeed the case by utilizing the
renormalization group equations.

The symmetry of the theory becomes evident by investigating the
bare action.
Under the gauge transformation (8), the bare action (14) changes as,
$$\eqalign{
\delta S ^0 = - \int d^D x \sqrt{\hat g} {\mu ^{\epsilon} \over G}
\{&{1\over 2}\{\epsilon L
- AG -2(D-1)(({\partial L\over{\partial \psi}})^2
-({\partial L\over{\partial \varphi _i}})^2)\}\tilde{R} \cr
&-{1\over {4}}\{\epsilon - 4(D-1) {\partial ^2 L\over{\partial \psi ^2}}\}
\partial _{\mu} \psi \partial ^{\mu} \psi \cr
&+{1\over {4}}\{\epsilon + 4(D-1)
{\partial ^2 L\over{\partial \varphi _j^2}}\}
\partial _{\mu} \varphi _i \partial ^{\mu} \varphi _i \}
\delta \rho ,
}\eqno\eq$$
where $\delta \rho = {2\over D} \partial _{\mu} \epsilon ^{\mu}$.
Note that this is nothing but the trace anomaly $({T^{\rho}}_{\rho})$
of the theory with respect to the reference (background)
metric $(\delta \hat{g}_{\mu\nu} = - \hat{g}_{\mu\nu} \delta \rho)$.
The only difference from the classical trace anomaly (10) comes from the
one loop counter term.
Since the second and third terms in (21) vanish identically, we only need to
consider the first term. If we substitute the explicit parametrization
of $L$, the coefficient of $\tilde{R}$ is,
$$\eqalign{
{\mu ^{\epsilon}\over G}\{
&\epsilon -GA -2(D-1)a^2
+ \{\epsilon -8(D-1)b\} a\psi \cr
&+ \{\epsilon -8(D-1)b\} b\psi ^2
- \{\epsilon -8(D-1)d\} d\varphi _i^2 \} \cr
={\mu ^{\epsilon}\over G} \{&\epsilon -GA -2(D-1)a^2\} .\cr
}\eqno\eq$$
Here again only the first term is nontrivial.
In order to ensure the gauge invariance of the theory, we have
to make sure that this quantity vanishes. In fact it
vanishes for the classical conformal Einstein gravity. So let
us pick a point in the weak coupling region of
$G$ such
that the trace anomaly vanishes there.
Then we consider the renormalization group trajectory
which passes through this point. The Einstein gravity
lies on this trajectory in the weak coupling limit.
By using the $\beta$ functions which are obtained in Sec. 3, we
can prove the invariance of the trace anomaly under the
renormalization group:
$$\eqalign{
\mu{\partial \over{\partial \mu}}
& \{{\mu ^{\epsilon}\over G} (\epsilon -AG -2(D-1)a^2 )\} \cr
=& 0 .
}\eqno\eq$$
Therefore the trace anomaly vanishes along the whole
renormalization group trajectory, if it vanishes in the
very weak coupling region.
Recall that the bare action possesses manifest volume preserving
diffeomorphism invariance. This symmetry is promoted to general
covariance by further demanding conformal invariance since we can replace
$\tilde{g}_{\mu\nu} \rightarrow g_{\mu\nu}$ in this case.
By this replacement, a manifestly volume preserving diffeomorphism
invariant action becomes a manifestly generally covariant action.
{}From these considerations, we conclude that the general covariance
of the theory is maintained along the renormalization group
trajectory which leads to the conformally coupled
Einstein gravity in the weak coupling limit.

This conclusion stems from the two points.
The first point is that the requirement of
general covariance is
equivalent to the vanishing of the trace anomaly
in our formulation.
The second point is the nonrenormalization theorem of the trace anomaly.

The nonrenormalization of the trace anomaly follows from the manifest
general covariance with respect to the reference metric in our
formulation. Therefore the reference metric is not renormalized.
The trace anomaly with respect to the reference metric
is defined as
$$
{T^\rho}_{\rho} = \hat{g}_{\mu\nu}{\delta S ^0\over{\delta \hat{g}_{\mu\nu}}},
\eqno\eq$$
where $S^0$ is the bare action and hence ${T^{\rho}}_{\rho}$ is a finite
operator.
It is not renormalized since
$$\eqalign{
\mu{\partial \over {\partial \mu}}{T^{\rho}}_{\rho}
&= \mu{\partial \over {\partial \mu}}
(\hat{g}_{\rho\sigma}{\delta S ^0 \over{\delta \hat{g}_{\rho\sigma}}})\cr
&=0 .
}\eqno\eq$$
%In fact nonrenormalization of trace anomaly
%holds in a very general context[].

Let us consider more general models defined by the following action
to see how these considerations
generalize:
$$
\int d^D x \sqrt{\hat g} {\mu ^\epsilon \over G} \{ L(X^i)\tilde{R}
+ {1\over 2} \tilde{g}^{\mu\nu} G_{ij} (X^k)\partial _{\mu} X^i
\partial _{\nu} X^j \} .
\eqno\eq$$
We would like to impose the following symmetry analogous to
(8),
$$\eqalign{\delta \tilde{g}^{\mu\nu}&=
\partial _{\mu} \epsilon ^{\rho}\tilde{g} _{\rho\nu}
+\tilde{g} _{\mu\rho} \partial _{\nu}\epsilon ^{\rho}
+\epsilon ^{\rho}\partial _{\rho}\tilde{g} _{\mu\nu}
-{2\over D}\partial _{\rho} \epsilon ^{\rho} \tilde{g}_{\mu\nu} ,\cr
\delta X^i &= \epsilon ^{\rho} \partial _{\rho} X^i
- (D-1) G^{ij} {\partial L\over {\partial X^j}}{2\over D}\partial _{\rho}
\epsilon ^{\rho} .
}\eqno\eq$$
To be precise we impose this symmetry on the bare theory as
a bare symmetry.
At the tree level, we can adopt the gauge fixing condition of (11).
The ghost action is also the same form with (12).
Note that this is nothing but the well known nonlinear sigma models
except the gauge fixing and the ghost terms (BRS trivial sector).
For the BRS nontrivial sector we can employ the knowledge of the
nonlinear sigma models[14][15][16].
In the background gauge we need not consider
the renormalization of the BRS trivial sector to the one loop level.
However beyond one loop
we also need to investigate the BRS trivial sector.
This problem is investigated in the next section.
By this way the one loop counter terms can be evaluated to be
$$
- \{ {{26-N}\over {24\pi \epsilon }}\}  \tilde{R}
+ {1\over{4\pi\epsilon}}\nabla ^i \partial _i L\tilde{R}
+{1\over {4\pi\epsilon}} R_{ij} \partial _\mu X^i
\partial _\nu X^j \tilde{g}^{\mu\nu} ,
\eqno\eq$$
where $N$ is the dimension of the `target' space.

Here we cite the remarkable connection between the trace anomaly
of the bare action and the $\beta$ functions of the theory[16].
Let us denote the couplings and the operators of the theory by
$\{\lambda _i \}$ and $\{\Lambda _i \}$.
In our case
they are $\{ L,G_{ij}\}$ and $\{\tilde{R} , \partial _{\mu} X^i
\partial _{\nu} X^j \tilde{g}^{\mu\nu}\}$.
The bare couplings are
$$
\lambda ^0_i = \mu ^{\epsilon} \{ \lambda _i + \sum _{\nu=1}^{\infty}
{a^{\nu}_i (\lambda _j )\over \epsilon ^{\nu}}\} .
\eqno\eq$$
The $\beta$ functions follow as
$$
\beta _{\lambda _i} = -\epsilon \lambda _i - a^1_i
+ \lambda _j {\partial \over \lambda _j} a^1_i .
\eqno\eq$$
The bare action of the theory is $S_0 =  \lambda ^0_i \Lambda ^0_i$.
The renormalized operator $\Lambda _i$ is defined as
$$
\Lambda _i = {\partial \over \partial \lambda _i} S_0 .
\eqno\eq$$
In conformally coupled Einstein gravity, there is no
operator renormalization to the one loop order.
The trace anomaly of the bare action $T^{\rho}_{\rho}$ is a
finite quantity and hence should be expressed by the
linear combinations of the renormalized operators with finite
coefficients. In fact these trace anomaly coefficients
can be related to the $\beta$ functions of the theory
to all orders.

%We recall that the conformal factor $e^{{\epsilon\over 2} \phi}$
%acts as a coupling constant.
%Hence the one loop amplitude is conformally invariant except
%terms which contain $\nabla ^\mu \partial _\mu \phi$ like
%${1\over{4\pi\epsilon}}\nabla ^i\partial _i L\tilde{R}$. However the
%one loop counter terms depend on the conformal factor
%just like the tree amplitude $(e^{-{\epsilon\over 2} \phi})$.
%Hence the one loop subtraction
%introduces the conformal
%anomaly proportional to the $\beta$ functions after cancelling the
%$1\over \epsilon$ poles.

%Including the tree level conformal anomaly, the trace anomaly
%of the effective action is,
In a general case, the trace anomaly becomes
$$\eqalign{
{\mu ^{\epsilon} \over G}
\{\epsilon  L - {{26-N}\over {24\pi}}G
+ {G\over{4\pi}} \nabla ^i \partial _i L + 2 (D-1)
\partial ^i L \partial _i L \}&\tilde{R} \cr
+{\mu ^{\epsilon}\over 2G}
\{\epsilon  G_{ij} + {G\over{2\pi}} R_{ij}
+4 (D-1) \nabla _i \partial _j L \}&\partial _{\mu}X ^i
\partial _{\nu} X ^j \tilde{g}^{\mu\nu} .
}\eqno\eq$$
%where $N$ is the dimension of the `target space'.
%In the background gauge we maintain explicit volume preserving
%diffeomorphism invariance. If the trace anomaly vanishes in addition,
%the effective action becomes invariant under the gauge
%transformation (27).
As we have explained in the beginning of this section,
the one loop counter terms can be chosen to preserve
the volume preserving diffeomorphism invariance
in general cases since the gauge invariance needs to be
only `softly' broken at $O(\epsilon )$ by the tree action.
The volume preserving diffeomorphism invariance of the bare
action can be promoted to full general covariance by
imposing conformal invariance.
In this way we construct the general covariant theory up to the
one loop level.

%The bare action is chosen to be invariant under the volume preserving
%diffeomorphism. It becomes generally covariant if the trace anomaly
%vanishes in addition.
%In this way we see the equivalence
%of the general covariance and the conformal invariance in a general
%context.
%
Now we consider higher loop amplitudes.
In order to renormalize the
two loop amplitudes, we need to subtract all subdivergences.
If we do so, all divergences are guaranteed to be local.
This point will be investigated in the next section.
%Since we maintain the explicit volume preserving diffeomorphism
%invariance in the background gauge, these divergences can be renormalized
%within the Lagrangian we have considered.
%The two loop amplitude itself is conformal invariant since
%it can be obtained from the lower loop amplitudes by
%contracting two fields or by the counter term insertion.
%The conformal invariance of the bare action
%in nonlinear sigma models is shown to be related
%to the $\beta$ functions of the theory to all orders[16].
The two loop conformal
anomaly is $O(G) \sim O(\epsilon )$ around the ultraviolet
fixed point which we are most interested in. Recall that we have
balanced the tree and the one loop conformal anomaly of $O(1)$
around the ultraviolet fixed point. So we can cancel the two loop
trace anomaly by slightly deforming $L$ and $G_{ij}$ by $O(\epsilon ^2 )$.
The two loop divergences in the theory
with the symmetry breaking of this magnitude
should be subtracted again by the counter terms which is invariant
under the volume preserving diffeomorphism because at most ${1\over \epsilon
^2}$
poles can appear at the two loop level.
At still higher orders, we repeat finer and finer deformations of
$L$ and $G_{ij}$. This procedure defines the $2+\epsilon$ expansion
of general quantum gravity to all orders.
The argument for the two loop case can be repeated inductively
for higher loop cases.
We hope a rigorous proof for the renormalizability
of $2+\epsilon$ dimensional quantum gravity
can be constructed along this line of arguments.

One of the technical subtleties of the renormalization program
concerns the BRS trivial sector. However the renormalization
of this sector must be done away with by a judicious choice
of gauge since there are infinite degrees of freedom
in gauge fixing. This point will be underscored by a concrete
calculation in the next section.
Therefore we expect no difficulty to renormalize this sector.
%if we can renormalize BRS nontrivial sector.
%If we accept such a hypothesis, the proof of the renormalizability
%of the theory to all orders follows from the power counting and
%the manifest volume preserving diffeomorphism invariance
%in the background gauge.

In conclusion we claim that the following trinity holds
in our formulation of quantum gravity near two dimensions:

The theory is generally covariant.

The theory is conformally invariant with respect to the reference metric
$\hat{g}_{\mu\nu}$.

The theory is on the renormalization group trajectory which leads to the
Einstein gravity in the weak coupling limit.

\vskip .10in

\centerline{{\bf 5.  Renormalization of quantum fields}}

\vskip .10in

In the background gauge we decompose the fields into the classical
(background) fields and the quantum fields. We functionally integrate
the quantum fields. The background fields may be viewed as the external
fields and the quantum fields as the internal fields.
It is at the heart of the BPHZ renormalization scheme that we subtract
all subdivergences to make the remaining overall divergence of a particular
diagram local.
We need to renormalize the quantum fields in general
to subtract subdivergences. In nongauge theories there is little difference
between the classical and quantum fields and the
renormalization of the quantum fields
is the same with that of the classical fields.
However in gauge theories the renormalization of the quantum fields
is in general different from that of the background fields
due to the gauge fixing. However this is essentially a
technical problem and we expect no difficulty in the
renormalization of the quantum fields if we can renormalize
the effective action of the background fields.
Since there are infinite degrees of freedom to choose a gauge,
it must be possible to choose a gauge in which there is no
need to renormalize the quantum fields. So formally we can argue that
we need not worry about the renormalization of the quantum fields.

Let us try to renormalize the theory to the two loop level
in the background gauge.
In this case we need to subtract all one loop subdivergences.
All such subdivergences can be viewed as the quantum two
point functions.
So we only need to make the quantum
two point functions finite to the one loop order
in order to ensure the two loop renormalizability of the
theory in the background gauge.
Since we have already made the effective action for the background fields
finite to the one loop order, we expect that
the quantum two point functions can be renormalized to the one loop order
also.

Since we are interested in the renormalization of the quantum fields,
we can use the flat background metric $({\hat g}_{\mu\nu} =
\eta _{\mu\nu} )$. The background dependence can be recovered by
the manifest covariance with respect to the background metric.
We adopt the following gauge fixing term
$$
{1\over 2} L (\partial ^{\mu} (h_{\mu\nu} + \lambda h_{\mu\rho}
h^{\rho}_{\nu} ) - {\partial _{\nu} L \over L} )^2 ,
\eqno\eq$$
where we introduce a gauge fixing parameter $\lambda$.
The ghost sector is
$$\eqalign{\partial _{\mu} \bar \eta _{\nu} \partial ^{\mu}\eta ^{\nu}
&+\partial _{\mu} \bar \eta _{\nu}
\partial _{\rho} h^{\mu\nu} \eta ^{\rho} \cr
+({1\over 2}+\lambda )
\partial _{\mu} \bar \eta ^{\nu} \partial ^{\mu}\eta ^{\rho}
h_{\rho\nu}
&-({1\over 2} -\lambda )
\partial _{\mu} \bar \eta ^{\nu} \partial ^{\rho}\eta ^{\mu}
h_{\rho\nu} \cr
-({1\over 2}-\lambda )
\partial ^{\mu} \bar \eta _{\nu} \partial ^{\rho}\eta ^{\nu}
h_{\mu\rho}
&+({1\over 2} +\lambda )
\partial ^{\mu} \bar \eta ^{\nu} \partial _{\nu}\eta ^{\rho}
h_{\mu\rho} \cr
-2\lambda \partial ^{\mu} \bar \eta ^{\nu} \partial _{\rho}\eta ^{\rho}
h_{\mu\nu} + \cdots .
}\eqno\eq$$

First we calculate the two point functions of $\psi$ and $\varphi _i$
fields. Just like the background gauge calculation,
we find no divergences of $\partial _{\mu} \psi \partial ^{\mu} \psi$
and $\partial _{\mu} \varphi _i \partial ^{\mu} \varphi _i$ type.
Next we consider the two point functions of $h_{\mu\nu}$.
There are only two types of them $\partial _{\rho} h_{\mu\nu}
\partial ^{\rho} h^{\mu\nu}$ and $\partial _{\mu} h^{\mu\rho}
\partial ^{\nu} h_{\nu\rho}$. The particular combination of them
appears in the quadratic part of $\tilde{R}$ as
$$
\tilde{R}_2 =
{1\over 4} (\partial _{\rho} h_{\mu\nu}
\partial ^{\rho} h^{\mu\nu} )
-{1\over 2} (\partial _{\mu} h^{\mu\rho}
\partial ^{\nu} h_{\nu\rho}) .
\eqno\eq$$
$\psi$ and $\varphi _i$ fields give the standard divergence of
$$
- {(1+c)\over {24\pi\epsilon}} \tilde{R}_2 .
\eqno\eq$$
The ghost contribution in this gauge is found to be
$$
-({1+12\lambda +4\lambda ^2\over 24\pi\epsilon}) \tilde{R}_2
+{\lambda ^2 \over 2\pi\epsilon }
\partial _{\mu} h^{\mu\rho}
\partial ^{\nu} h_{\nu\rho} .
\eqno\eq$$

When we evaluate $h_{\mu\nu}$ field contribution, it is convenient
to use the background field method.
Namely we expand $\tilde{g}_{\mu\nu} =(e^{h})_{\mu\nu}$
around the expectation value
of $h_{\mu\nu}=\hat{h}_{\mu\nu} +h'_{\mu\nu}$:
$$\eqalign{
\tilde{g}_{\mu\nu} &=(e^{\hat h +h'})_{\mu\nu} \cr
&=  (e^{\hat h (1-{G\over 6\pi \epsilon L})})_{\mu\rho}
(e^{\tilde{h}})^{\rho}_{\nu} .
}\eqno\eq$$
Here we have to be careful about the one particle irreducibility
of the fields which are related nonlinearly.
If we parametrize ${\tilde{h}^{\mu}}_{\nu}$ as in [7][8]
$$
{\tilde{h}^{\mu}}_{\nu} = \hat g ^{\mu\rho} H_{\rho\nu}
-{1\over D} {\delta ^{\mu}}_{\nu} \hat g ^{\rho\sigma}
H_{\rho\sigma},
\eqno\eq$$
then $h' _{\mu\nu}$ and $H_{\mu\nu}$ are related nonlinearly
$$\eqalign{
h' _{\mu\nu}& =
H_{\mu\nu} -{1\over 2} (H{\hat h}+{\hat h}H)_{\mu\nu}
+{1\over 2} \eta _{\mu\nu} \hat h ^{\rho\sigma} H_{\rho\sigma} \cr
&+ ({1\over 12}\hat h H^2 -{1\over6}H\hat h H
+{1\over 12}H^2 \hat h +{G\over 6\pi\epsilon L}\hat h )_{\mu\nu}.
}\eqno\eq$$
The one particle irreducibility of $h' _{\mu\nu}$ and
$H_{\mu\nu}$ $(< h' _{\mu\nu}>=
<H_{\mu\nu}>=0)$ can be made consistent only after the subtraction
(the last term on the right hand side).
This is the reason that the such a singular term appears in (38).
However this term can be cancelled by the renormalization
of the quantum field.

Now we can expand the action using our standard parametrization
$\tilde g _{\mu\nu} = (\hat g )_{\mu\rho} (e^{\tilde{h}} )^{\rho}_{\nu}$
just like the background gauge calculation.
The gauge fixing term can be expanded around the mean field and
it is found to be
$$
{1\over 2}
(\partial ^{\mu}
(\hat h +H +(\lambda -{1\over 2})(H\hat h +\hat h H)
+{1\over 2}\eta \hat h _{\rho\sigma} H^{\rho\sigma} )
_{\mu\nu})^2.
\eqno\eq$$
In this way we find the contribution of $h_{\mu\nu}$ field as
$$
{27+60\lambda +4\lambda ^2\over 24\pi\epsilon }{\tilde R}_2
-{\lambda ^2\over 2\pi \epsilon}
\partial _{\mu} h^{\mu\rho}
\partial ^{\nu} h_{\nu\rho}
+{1\over 3\pi \epsilon}{\tilde R}_2 ,
\eqno\eq$$
where the last term is due to the change of the variables
we have just explained.

In this way the total divergence is found to be
$$
{25-c +48\lambda +8 \over 24\pi \epsilon} {\tilde R}_2 ~.
\eqno\eq$$
Just like the Yang-Mills theory, the divergence of the two
point functions of $h_{\mu\nu}$ is of the form $\tilde{R}_2$.
The ghost contribution is essential for this result.
We also conclude that there is no renormalization of the
gauge fixing term for $h_{\mu\nu}$ fields.
Since the first term (25-c) can be absorbed in the gravitational coupling
constant renormalization, we need the wave function renormalization
as
$$
h^0_{\mu\nu} = h_{\mu\nu}
(1-{1+6\lambda \over 6\pi\epsilon}{G\over L}) .
\eqno\eq$$

Lastly we evaluate the divergent part of the
ghost two point function to be
$$
{G\over 12\pi \epsilon L} \partial _\mu {\bar \eta} _{\nu}
\partial ^{\mu} \eta ^{\nu} .
\eqno\eq$$
It is clear that this divergence can be subtracted by the
ghost wave function renormalization.

Furthermore we can check the consistency of the result
by calculating three point functions such as
$\partial _{\mu} \psi \partial _{\nu} \psi
h^{\mu\nu}$.
The one loop divergence of this type is found to be
$$
{1+6\lambda \over 12\pi\epsilon L}
\partial _{\mu} \psi \partial _{\nu} \psi
h^{\mu\nu} ,
\eqno\eq$$
which is consistent with the wave function renormalization of
$h_{\mu\nu}$.

In fact there are other types of the quantum two point functions
which contain the derivatives of $L$ such as
${\partial _{\mu} L\over L} \partial _{\nu} h^{\mu\nu}$.
Such terms are $O(a \sim \sqrt{\epsilon})$
in generic cases.
They have to be dealt with since ${1\over \epsilon}$ pole
can arise in the one loop integration.
It is straightforward to calculate these divergences
and we expect no difficulty to subtract them
since the tree level symmetry breaking is $O(\epsilon )$.
However they become $O(\epsilon )$ at the
short distance fixed point which we are most interested in
since $a$ vanishes there. So there is no subdivergence
of this type at short distance and we are content not to
discuss them in this paper.
%in conformally coupled (and in general) Einstein gravity.
%Since this term has to be
%contracted with the same type of the vertices $(\psi h)$ to become
%a two loop diagram, such a diagram is $O(\epsilon )$. So there occurs no
%subdivergence in this class of diagrams. Hence we need not subtract
%such quantum two point functions.

Note that the wave function renormalization for $h_{\mu\nu}$ field
is not necessary if we choose $\lambda =-{1\over 6}$. This result
underscores our expectation that the renormalization
of the quantum fields can be done away with by a judicious gauge fixing.
In this section we have performed all the necessary wave function
renormalization of the quantum fields to ensure that the two
loop counter terms are local at short distance.
Due to the manifest volume preserving
diffeomorphism invariance
in the background gauge, these divergences can be
renormalized with our action.
So we are certain that the theory can be renormalized to the two
loop order and hope to report the results of such a calculation
in the near future.

\vskip .10in

\centerline{{\bf 6.  Renormalization of Composite Operators}}

\vskip .10in

In this section we consider the renormalization of the
composite operators such as the cosmological constant operator.
In the classical theory the cosmological constant operator is
$$\eqalign{
\int d^D x \sqrt{g} &= \int d^D x\sqrt{\hat g} (1+{1\over 2}
\sqrt{{\epsilon \over 2(D-1)}}\psi )^{2D\over \epsilon} \cr
&= \int d^D x\sqrt{\hat g} exp(\sqrt{2\over \epsilon}\psi
- {1\over 4}\psi ^2 +\cdots ) .
}\eqno\eq$$
In quantum theory, this operator is renormalized.
We consider the infinitesimal perturbation of the theory
by the cosmological constant operator.
The renormalized cosmological constant operator can be
determined by requiring that the operator is invariant
under the gauge transformation (8).
Since the gravitational dressing of the spinless operators
involve only $\psi$ field, we require the invariance with respect
to the following transformation
%at the short distance fixed point:
$$\eqalign{
\delta \psi &= (a' + {\epsilon \over 4} \psi )\delta \rho ,\cr
\delta {\hat g}_{\mu\nu} &= -{\hat g}_{\mu\nu} \delta \rho ,
}\eqno\eq$$
where $a' = (D-1)a$.
This requirement is equivalent
to impose the background independence of the theory
including the cosmological constant operator.

Let the cosmological constant operator to be
$\int d^D x \sqrt{\hat g} \Lambda (\psi )$. We parametrize
$\Lambda (\psi ) = exp(\alpha \psi + {\beta \over 2} \psi ^2 +\cdots )$.
If we vary the background metric as in (48),
this operator varies due to the one loop quantum effect as
$$\delta \Lambda (\psi ) =
{G \over 8\pi} {\partial ^2 \over \partial \psi ^2} \Lambda (\psi )
\delta \rho . \eqno\eq$$
The variation due to $\psi$ field transformation as in (48) is
$$
\delta \Lambda (\psi ) = (a' + {\epsilon \over 4} \psi ){\partial \over
\partial \psi}
\Lambda (\psi ) \delta \rho .
\eqno\eq$$
The sum of the above must cancel the variation of
$\delta (\sqrt{\hat g} )\Lambda = -{D\over 2}
\sqrt{\hat g} \Lambda \delta \rho$ .
% caused by the change of $\sqrt{\hat g}$.

The coefficients $\alpha ,\beta ,\cdots $ are determined in this way as
$$\eqalign{
\alpha &= {4\pi a\over G} (-1 \pm \sqrt{1+{G\over 2\pi a^2}}) ,\cr
\beta &= -{\epsilon \alpha \over 4a + {G\alpha \over \pi}} ,\cr
&\cdots .
}\eqno\eq$$
We choose the $+$ sign out of the two possible
branches in the above expression since it agrees with
the classical expression in the weak coupling limit.
This strategy determines the renormalized cosmological constant
operator around the ultraviolet fixed point:
%to the leading order of $\epsilon$ to be
$$
\int d^D x \sqrt{\hat g}
exp({Q \over \sqrt{\epsilon}} (1+{5\over 16} \epsilon )
\psi -{Q^2 \over 16}\psi ^2 + \cdots ) ,
\eqno\eq$$
%where ${\alpha ^2 G^* \over 8\pi}=1$.
%So $\alpha = \pm {Q \over \sqrt{\epsilon}}$
where $Q^2 = {25-c \over 3}$.
%For $\psi >0$, $\Lambda (\psi )$ increases while $\Lambda (-\psi )$ decreases
%if we increase the magnitude of $\psi$. In this sense the former is a relevant
%operator and the latter is an irrelevant operator.
%Therefore we pick the former as the cosmological constant operator.

Recall that at the short distance fixed point, the theory is
invariant under $\psi \rightarrow -\psi$.
%This symmetry is reminiscent to the target space duality
%in string theory[17].
%So we have added two possible operators with the equal weights
%to respect this invariance.
In this context we point out the analogy with the Ising model which also
possesses the $Z_2$ discrete symmetry. The cosmological constant operator
is analogous to the spin operator since it is not invariant under the $Z_2$
symmetry. In fact $\Lambda (\psi )  \rightarrow \Lambda (-\psi )$ under
this transformation.
Physically speaking, we are considering the fluctuations
of the metric around the vanishing
expectation value. Under such a circumstance, the cosmological constant
operator may fluctuate into the negative (or complex) value.

In quantum gravity the cosmological constant operator may serve the order
parameter just like the spin operator in the Ising model.
We can think of two distinctive phases of quantum gravity which
possess vanishing and nonvanishing expectation values of
the cosmological constant operator. Obviously we are in the
phase which possesses the nonvanishing expectation value of the
cosmological constant operator since our Universe is certainly
large. In the other phase the macroscopic spacetime does not
exist and it might be a topological phase.
In a manifestly covariant subtraction scheme (1),
%the effective
%Liouville theory is obtained as in (1). Note that at the short distance
%fixed point,
the sign of the kinetic term of the conformal mode
$(Q^2_{eff})$ flips. In this sense, we may
rotate $\psi \rightarrow i\psi$ in the `topological' phase.
In doing so, the cosmological constant operator becomes a
complex operator
as it is the case in two dimensional quantum gravity with
the central charge $1<c<25$. Then the analogy with a
spin model can be drawn and the vanishing expectation value of the cosmological
constant operator is conceivable.

One of the great mysteries in Nature is the cosmological
constant problem. We have to explain why the observed cosmological
constant is so small compared to the planck scale (or any other
subatomic scale). One of the possible explanations is to
invent a symmetry. We observe that this $Z_2$ symmetry can
prohibit the cosmological constant. Therefore we speculate that
the resolution of the cosmological constant problem
in Nature might be
due to such a discrete symmetry.

At the fixed point, the coefficient of $\tilde{R}$ is
${1\over G^*} (1+{\epsilon\over 8(D-1)}\psi ^2 )$. This acts
as the effective inverse coupling at the scale set by $\psi$.
Let $G=G^* -\delta G$, then the $\beta _G$ function in (17) becomes
$\mu {\partial \over \partial \mu}
\delta G = -\epsilon  \delta G$ around the fixed point.
We can identify ${\delta G \over G^*} = {\epsilon\over 8(D-1)}\psi ^2$.
Since $\mu {\partial \over \partial \mu} \psi = -{\epsilon \over 2}\psi$
around the fixed point, the right hand side satisfies the same
renormalization group equation with the left hand side.
So such an identification is consistent with the renormalization group
considerations in sec. 3.

In the classical limit, the coefficient of $\tilde{R}$ is
%${1\over G}(1+{1\over 2}\sqrt{\epsilon \over 2(D-1)}\psi )^2$.
${1\over G} e^{-{\epsilon \over 2}\phi}$ as in (4).
This is the effective inverse gravitational coupling at long distance.
%If we scale the cosmological constant operator (44) by $\lambda ^{2D\over
%\epsilon}$, the
%inverse gravitational coupling scales as $\lambda ^2$. This scaling
%behavior displays nothing
%but the canonical dimensions of these couplings.
The cosmological constant operator in the classical limit is
$\Lambda = e^{-{D\over 2}\phi}$.
%The classical scaling dimension of the inverse
%gravitational coupling constant relative to the cosmological constant
%operator is
%$$
%{d \over d log \Lambda} {1\over G} = {\epsilon \over D} {1\over G} .
%\eqno\eq$$
The classical scaling relation is
$$
\Lambda ^{\epsilon} \sim ({1\over G}) ^D .
\eqno\eq$$

Now let us consider how this scaling behavior changes at short
distance.
%Around the short distance fixed point, the cosmological
%constant operator
%$\Lambda (\psi )$
%behaves as ${Q^2 \over \epsilon } \psi ^2 + O(\psi ^4)$.
%Here we have discarded the constant term.
%So $\delta \Lambda$
%scales the same way with $\delta G$ if we change the scale
%of $\psi$. In this sense, we say that the cosmological constant
%operator and the inverse gravitational coupling possess the same
%scaling dimensions at the short distance limit.
%So the quantum effect is found to work in the right direction
%to resolve the naturalness problem of the cosmological constant
%of our Universe which has bee found to be also the case in the minimally
%coupled Einstein gravity[8]. Unfortunately this effect alone
%is nowhere
%sufficient and fine tuning of the theory is necessary to realize the
%large universe like our own.
Around the short distance fixed point, $\delta \Lambda (\psi ) \sim
{Q\over \sqrt{\epsilon}} \psi$.
We find the following scaling relation between $\delta ({1\over G})$ and
$\delta \Lambda$.
$$
(\delta \Lambda ) ^2 \sim \delta ({1\over G }) .
\eqno\eq$$
Remarkably the cosmological constant operator is no longer the
most relevant operator at the short distance fixed point.
The quantum renormalization effect also works in the right direction
to resolve the cosmological constant problem.
%Hence
%$$
%{d \over d log \Lambda} {1\over G} =
%{Q\over 32\pi}\sqrt{\epsilon} \psi .
%\eqno\eq$$
%Obviously it does not scale in a simple manner.
%However the appearance of $\sqrt{\epsilon}$ factor
%is analogous to the minimal coupling case[8].

\vskip .10in

\centerline{{\bf 7.  Quantum Cosmology in $2+\epsilon$ Dimensions}}

\vskip .10in

In this section, we discuss the cosmological implications of
this model. Let us consider the Robertson-Walker spacetime
in $D$ dimensions
$$\eqalign{
ds^2 &= -dt^2 + r(t)^2  \tilde{g}_{ij} dx^i dx^j ,\cr
\tilde{g}_{ij} &= \delta _{ij} +{kx_ix_j \over {1-k|\vec x |^2}} .
}\eqno\eq$$
The Ricci curvature is
$$\eqalign{
R_{tt} &= {\ddot r\over r}(D-1) ,\cr
R_{ij} &= -(r\ddot r +\epsilon (\dot r )^2 +\epsilon k )\tilde g _{ij} ,
}\eqno\eq$$
where $\dot r = {d\over dt}r$.
Let the energy momentum tensor of the matter as
$$
T_{\mu\nu} = pg_{\mu\nu} + (p+\rho )U_{\mu}U_{\nu} ,
\eqno\eq$$
with $U^t = 1$ and $U^i = 0$.
The equation of the energy conservation is
$$
{d\over dr}(\rho r^{D-1} ) = -(D-1) p r^{D-2} .
\eqno\eq$$
For the conformal matter where $p={\rho\over D-1}$, we find
$\rho \sim r^{-D}$.
The Einstein's field equation is
$$
R_{\mu\nu} -{1\over 2} g_{\mu\nu} R = -8\pi G T_{\mu\nu} .
\eqno\eq$$

Alternatively we may use the Wheeler-DeWitt equation[17], which
can be obtained from the action
$$
S = 2\int _{\partial M} d^{D-1} x \sqrt{h} K
+ \int _M d^D x \sqrt{g} R + (matter) .
\eqno\eq$$
The second term is integrated over spacetime and the first over
its boundary. $K$ is the trace of the extrinsic curvature $K_{ij}$
of the boundary.
By taking the variation with respect to the lapse, we obtain the
Hamiltonian constraint
$$
\sqrt{h}(K^2 -K_{ij}K^{ij} - R^{(D-1)} -16\pi G \rho ) =0 .
\eqno\eq$$
For the Robertson-Walker metric, we find:
$$\eqalign{
K_{ij} &=r\dot r \delta _{ij} ,\cr
K&={\dot r \over r} (D-1) ,\cr
R^{(D-1)} &= -\epsilon (D-1) {k\over r^2} .
}\eqno\eq$$
Either from the Einstein equation or from the Wheeler-DeWitt
equation, we obtain
$$
\dot r ^2 + k = {16\pi G\over \epsilon (D-1)}\rho r^2 .
\eqno\eq$$
The right hand side of this equation possesses the two dimensional limit if
$G \sim O(\epsilon )$ and we denote it by
${8(\Delta _0 -1)\over |Q^2 |}$.

We point out the similarity of this equation with $k=1$
(closed universe) to the Wheeler-
DeWitt equation of the Liouville theory.
This similarity is not surprising since the Liouville theory
can be obtained from the Einstein action in $2+\epsilon $
dimensions by taking $\epsilon \rightarrow 0$ limit as
it is explained in (1).
%$$\eqalign{
%{1\over G} &\int d^D x\sqrt{\hat g}e^{-{\epsilon\over 2}\phi}\{\tilde{R} -
%{\epsilon (D-1)\over 4}
%\partial _{\mu}\phi\partial _{\nu}\phi\tilde{g}^{\mu\nu}\} \cr
%\rightarrow & {Q^2 \over 32\pi}
%\int d^D x \sqrt{\hat g} (\partial _{\mu}\phi
%\partial ^{\mu}\phi +2\phi \hat R) ,
%}\eqno\eq$$
%where the identification ${-\epsilon \over G} = {Q^2 \over 8\pi}$
%is made.
Recall that the weak coupling region of $2+\epsilon $
dimensional quantum gravity is related to the Liouville theory
with $c > 25$.
The Wheeler-DeWitt equation in the minisuperspace approximation is[18]:
$$
[{Q^2\over 8}
(-({\partial \over ({Q\over 2})^2 \partial \phi} )^2 +1)
+\Delta _0 -1]\Psi =0 ,
\eqno\eq$$
where $\Psi = e^{Q(\beta +{Q\over 2})\phi} \sim
e^{\pm i\sqrt{2\Delta _0}|Q|\phi}$.
This solution in the minisuperspace approximation possesses the complex
Liouville exponents.  This wave function is (delta function)
normalizable with the standard $L^2$ norm:
$\| \Psi \| ^2 = \int _{-\infty}^{\infty} d\phi |\Psi (\phi )|^2$.

In order to represents a macroscopic universe like our own,
we need to form a wave packet with nontrivial matter contents:
$$
\int dp  a(p) e^{ip\phi} \Phi _{matter} (p,x_0)
\eqno\eq$$
where $\Phi _{matter} (p,x_0)$ is a time $(x_0)$ dependent matter
wave function and $a(p)$ is a certain weight to form a wave packet.
If we draw the analogy with the particle quantum mechanics,
we think of the following wave packet:
$$\eqalign{
&\int {dp\over \sqrt{2\pi}}  e^{-{1\over 2}(p-p_0 )^2} e^{ip\phi} e^{i{p^2
\over 2}x_0} \cr
\sim & {1\over (1+x_0^2)^{1\over 4}}
exp(-{1\over 2} {1\over 1+x_0^2}
(\phi -x_0 p_0)^2) ,
}\eqno\eq$$
where we have suppressed the phase factor.
Since $\phi$ is the scale factor of the universe, this wave
packet represents an expanding (or contracting) universe.

{}From these considerations it is clear that the classical solutions
of the Einstein theory correspond to the wave functions with
complex Liouville exponents. We emphasize that the states with
complex Liouville exponents (macroscopic states) must occur in the
theory in order to ensure the existence of spacetime itself.
It is certainly not a trivial requirement since we know that all
physical states possess real Liouville exponents in
two dimensional quantum gravity with $c \le 1$. In this sense
the existence of macroscopic spacetime itself can
serve as the order parameter in quantum gravity.
{}From the view point of the quantum gravity the $c=1$ barrier of
two dimensional quantum gravity is not a problem but a blessing
because macroscopic states can exist for $c > 1$. In fact
the weak coupling region of $2+\epsilon$ dimensional quantum
gravity is very similar to two dimensional quantum gravity with $c > 25$.

In the remaining part of this section we study the singularity
of the spacetime which we have to face at the beginning (or end) of the
Universe. Since we have constructed the consistent theory of
quantum gravity near two dimensions, we must be able to answer
this question. Furthermore our theory is still weak coupling
at the short distance fixed point since $G^* \sim \epsilon$.
So we must be able to answer this question by solving the
classical equations of the renormalized action:
$$\eqalign{
\int d^D x\sqrt{\hat g} {\mu ^{\epsilon}\over {G}} &\{ \tilde{R}
(1+ a \psi +{\epsilon \over{8(D-1)}}(\psi ^2 - \varphi _i^2 ))\cr
-&{1\over 2}\partial _{\mu}\psi\partial _{\nu}\psi \tilde{g}^{\mu\nu}
+{1\over 2}\partial _{\mu}\varphi _i\partial _{\nu}
\varphi _i \tilde{g}^{\mu\nu}\}.
}\eqno\eq$$
By taking the variations with respect to $h_{\mu\nu},\psi$ and
$\varphi _i$ around $\tilde{R} =0$, we obtain the field equations
$$\eqalign{
-&a\partial _{\mu} \partial _{\nu} \psi \cr
-&{\epsilon \over 8(D-1)} \partial _{\mu} \partial _{\nu} \psi ^2
+{1\over 2} \partial _{\mu} \psi \partial _{\nu} \psi \cr
+&{\epsilon \over 8(D-1)} \partial _{\mu} \partial _{\nu} \varphi _i^2
-{1\over 2} \partial _{\mu} \varphi _i \partial _{\nu} \varphi _i =0 ,\cr
&\partial _{\mu} \partial ^{\mu} \psi =0 ,\cr
&\partial _{\mu} \partial ^{\mu} \varphi _i =0 .
}\eqno\eq$$
The cosmological solution is
$$\eqalign{
\psi &= {1\over \sqrt{c}} \varphi _i = x_0 ,\cr
h_{\mu\nu} &= 0 .
}\eqno\eq$$

In the classical limit the conformal factor is
$e^{-\phi}\sim {x_0}^{4\over \epsilon}$.
By equating the line elements
$$\eqalign{
ds^2 &= e^{-\phi}(-dx_0^2 + \tilde{g}_{ij}dx^idx^j) \cr
&= -dt^2 + r^2\tilde{g}_{ij}dx^idx^j ,
}\eqno\eq$$
we find
$t \sim {x_0}^{D\over \epsilon}$
and $r \sim t^{2\over D}$.
Note that this solution is consistent with (63) if $k=0$ since we
have assumed $\tilde{R} =0$.
% However we expect that the difference
%of $k$ can be neglected at short distance.

As we trace the history back to the big bang, the Lagrangian itself
is renormalized.
%The remarkable point is that
This solution remains to be valid
not only in the classical limit but also throughout the renormalization
group trajectory up to the short distance fixed point where $a=0$.
At the short distance fixed point, the Lagrangian possesses the discrete
symmetry $(\psi \rightarrow -\psi,\varphi _i \rightarrow -\varphi _i )$
which may be called as `time reversal' symmetry. If we trace the history
of the universe before the big bang $(\psi <0)$, the universe looks like
just the time reversal of what happens for $\psi >0$. So eventually
it reaches the classical region again where $\psi ^2$ is very large and
the gravitational coupling is very small. Thus we conclude that the
universe bounces back from the big crunch due to the quantum effect
in quantum gravity near two dimensions.

Now we consider $k=\pm 1$ case at the short distance
fixed point where $a=0$.
By the way the equations of motion are identical to
those in the classical limit.
We consider symmetric solutions between $\psi$ and $\varphi _i$
just as $k=0$ case.
Then the only difference turns out to be the introduction of
mass terms for $\psi$ and $\varphi _i$ fields. We find that $\psi$ and
$\varphi _i$ fields are governed by the following effective
Lagrangian
$$
L = {1\over 2} {\dot \psi}^2 -{k\over 8} \epsilon ^2 \psi ^2 .
\eqno\eq$$
A point particle which moves according to this Lagrangian
represents the scale of the universe.
If $k=1$, the potential is harmonic and the universe cannot
expand forever. On the other hand, the $k=-1$ case corresponds
to the inverted harmonic potential and the universe expands
forever.
At short distance the difference of $k$ is irrelevant as we
expected.
%The short distance behavior of the $k=1$ case is identical
%to that of the $k=0$ case. In the $k=-1$ case the bounce back
%possibility before reaching $\psi =0$
%point exists.

If we assume that the universe is described by a wave function,
then it is not surprising to find that the spacetime singularity
is resolved quantum mechanically[17]. However
in quantum gravity near two dimensions
we can derive such a conclusion from the first principle
since a consistent theory
can be constructed in this case.

If the universe is open, the universe is symmetric around the
big bang and it expands in both the future and the past directions.
In such a situation the direction of time might be reversed before
the big bang.
On the other hand, the closed universes inevitably face the big crunch.
So we must conclude that the
closed universes near two dimensions cannot
escape the karma of infinite cycles of death and rebirth.
The only possibility to escape this infinite cycle
is presumably the quantum tunneling mechanism
(topology change).
% just light sudden enlightenment.

\vskip .10in

\centerline{{\bf 8.  Conclusions and Discussions}}

\vskip .10in

In this paper we have studied the conformally coupled Einstein gravity.
We have formulated quantum gravity in $2+\epsilon$ dimensions
in such a way to preserve the volume preserving diffeomorphism
invariance. We have proposed such a formulation in order to
keep track the dynamics of the conformal mode in a renormalizable
fashion. We have used the conformal Einstein gravity as a testing
ground to demonstrate the effectiveness of our formulation.

In such a formulation, the prescription to enforce the general covariance
is crucial. We have identified the gauge transformation which ensures the
general covariance of the theory.
In our formulation, this gauge invariance holds if the theory is conformally
invariant with respect to the background metric.
It is also shown that such a requirement is equivalent to choose theories
on the renormalization group trajectory which leads to the
Einstein gravity at long distance.

We also discussed the physical implications of the conformally coupled
Einstein gravity.
As it is explained in the introduction, the dynamics of the conformal mode
can be understood by the oversubtraction (1). In conformally couple
Einstein gravity,
there is no other factors which influence the dynamics of the conformal
mode unlike the minimal coupling case[8].
So the sign of the kinetic term of the conformal mode is expected to flip
at precisely the same point where $\beta _G = 0$. This is not so in the
minimal coupling case.
Our results in the formulation which avoids the oversubtraction is
consistent with such a picture.

Therefore the short distance fixed point we have studied in this paper
is not an ordinary second order phase transition point which appears
in the conventional field theory. At this point, the signature of the conformal
mode changes from spacelike to timelike.
Recall that the conformal mode is identified as the macroscopic `time'
in the wave packet construction (63).
Therefore we may say that the time is born in this phase transition.
The cosmological constant operator may serve the order parameter
as we have discussed in Sec. 5.

We have also discussed the implications of this model for
the spacetime singularity at the big bang or the end of the
blackhole evaporation. We have found that the universe bounces back
from the big crunch due to the quantum renormalization
effect. In this context we may wonder what happens to the Ricci
curvature singularity at short distance.
We expect that the spacetime becomes confomally invariant and
selfsimilar at the fixed point. Then the singularity may be
resolved in a simple scaling.
In order to answer this question, we need to renormalize such
generally covariant operators.
We expect that the physical operators are covariant
under the volume preserving diffeomorphism.
We presumably need to require that the physical operators
must scale under the conformal transformation.
It is not at all obvious that such operators can be
constructed and this problem requires more investigations.
%We believe that it is a very interesting problem
%to investigate.

We also hope that our approach sheds light on the
blackhole physics. The singularity which appears
in the final stage of blackholes is very similar
to the singularity we have discussed. So we expect that
the blackholes also bounce back in the final stage
into the original universe. However the riddles of
blackhole physics arise largely from the presence
of the event horizons. Therefore we need to investigate
the event horizons also. We expect to find no uncontrollable
divergences in physical quantities in this domain
since we have constructed a consistent theory.

As we have emphasized, our formulation can be viewed as
a Ginzburg-Landau type theory which is capable to study
the phase transitions in quantum gravity.
In the conventional approach, the metric is expanded
around the nonvanishing expectation value. In fact it is impossible
to expand around the vanishing expectation value of the
metric in a generally covariant formulation.
Our formulation allows us to do so by relaxing the manifest
symmetry. In fact the short distance fixed point we have found
can be interpreted as such a point. We may draw the analogy with the
nonlinear sigma models here again since
the order parameter also vanishes at the short distance fixed point.
%of the nonlinear sigma models is also a point
%at which .
We would like to extract more physical insights concerning the
phase transitions in quantum gravity using this tool.
We also believe that a systematic $2+\epsilon$ expansion
of quantum gravity is possible in our formulation.
We plan to report further progress on this project
in the near future.

%Here we make a conjecture $1$.
%All subdivergences of $O(\hbar ^n )$ which have to be subtracted
%to renormalize the theory to the (n+1)-th loop level can be
%subtracted by the bare action if the theory is renormalized
%to the n-th loop level.
\vskip .10in

{\bf Acknowledgements}

This work is supported in
part by Grant-in-Aide for Scientific Research from the Ministry
of Education, Science and Culture.

\endpage
\centerline{{\bf References}}

[1] S. Weinberg, in General Relativity, an Einstein Centenary Survey,

	   ~~~~~eds. S.W. Hawking and W. Israel (Cambridge University Press, 1979).

[2] R. Gastmans, R. Kallosh and C. Truffin, Nucl.Phys. {\bf B133} (1978) 417.

[3] S.M. Christensen and M.J. Duff, Phys. Lett. {\bf B79} (1978) 213.

[4] V.G. Knizhnik, A.M. Polyakov and A.A. Zamolochikov,

    ~~~~~Mod.Phys.Lett.{\bf A3} (1988) 819.

[5] H. Kawai and M. Ninomiya, Nucl. Phys. {\bf B336} (1990) 115.

[6] See for example, J. Ambjorn, S. Jain, J. Jurkiewicz and C.F. Kristjansen,

    ~~~~~Phys. Lett. {\bf B305} (1993) 208.

[7] H. Kawai, Y. Kitazawa and M. Ninomiya, Nucl. Phys.
{\bf 393} (1993) 280.

[8] H. Kawai, Y. Kitazawa and M. Ninomiya, Nucl. Phys.
{\bf 404} (1993) 684.

[9] I.R. Klebanov, I.I. Kogan and A.M. Polyakov, Phys. Rev. Lett. {\bf 71}
(1993) 3243.

[10] S. Kojima, N. Sakai and Y. Tanii, Phys. Lett. {\bf B322} (1994) 59;

     ~~~~~TIT-HEP 238 (1993).

%				Super Gravity in $2+\epsilon$ Dimensions, TIT-HEP 238 (1993).

[11] J. Nishimura, S. Tamura and A. Tsuchiya, UT-664,ICRR-Report-306-94-1

     ~~~~~and UT-Komaba/94-2.

%Scaling Dimensions of Manifestly Generally
%     Covariant Operators in Two-Dimensional Quantum Gravity,

[12] J. Distler and H. Kawai, Nucl. Phys. {\bf B321} (1989) 509.

[13] F. David, Mod. Phys. Lett. {\bf A3} (1988) 651.

[14] C.G. Callan, D. Friedan, E.J. Martinec and M.J. Perry,

     ~~~~~Nucl. Phys. {\bf B262} (1985) 593.

[15] E.S. Fradkin and A.A. Tseytlin, Nucl. Phys. {\bf B261}
(1985) 1.

[16] A.A. Tseytlin, Nucl. Phys. {\bf B294} (1987) 383.

%[17] K. Kikkawa and M. Yamasaki, Phys. Lett. {\bf B149}
%(1984) 357;
%
%     ~~~~~N. Sakai and I. Senda, Progr. Theo. Phys. {\bf 75} (1986) 692.
[17] J.B. Hartle and S.W. Hawking, Phys. Rev. {\bf D28} (1983) 2960.

[18] N. Seiberg, Prog. Theo. Phys. Suppl. {\bf 102} (1990) 319.

%[1] E. Br\'ezin, V. Kazakov; Phys. Lett {\bf B236} (1990), 914.
%D. Gross, A. Migdal; Phys. Rev. Lett. {\bf 64} (1990), 127.
%M. Douglas, S. Shenker; Nucl.Phys. {\bf B335} (1991), 589.

%[4] G. Moore, N. Seiberg, M. Staudacher; Nucl. Phys. {\bf B362} (1991), 665.

\end